\documentclass[]{jd04}    % Specifies the document style.
\usepackage[]{graphicx}

\Title[Asymptotic giant branch evolution and planetary nebulae]{
Asymptotic giant branch evolution and its impact on the chemical 
evolution of the Milky Way and the Magellanic Clouds} \Author{
Stanghellini$^1$, Letizia} 
\Address{$^1$National Optical Astronomy Observatory,
950 N. Cherry Avenue, Tucson, AZ  85719, USA
}{}

\ShortAuthors{Stanghellini}


\begin{document}
\maketitle

\begin{abstract}
The asymptotic giant branch (AGB) phase of stellar evolution is common to most stars of
low and intermediate mass. Most of the carbon and nitrogen
in the Universe is produced by AGB stars. The final fate of the AGB envelopes are represented by
planetary nebulae (PN). By studying PN abundances and compare them with the yields of stellar
evolution is possible to quantify carbon and nitrogen 
production, and to study cosmic recycling in galactic and Magellanic Cloud populations. 
In this paper we present the latest results in PN chemical abundance analysis
and their implication to the chemical evolution of the galaxy and the Magellanic
Clouds, with particular attention to carbon abundance, available only thanks to
ultraviolet spectroscopy.

\end{abstract}

\section{Introduction}
Low- and intermediate-mass stars (M$_{\rm MS} \sim 0.8 - 8$ M$_{\odot}$, where 
M$_{\rm MS}$ is the main sequence mass) and their 
ejecta are excellent probes of stellar populations in galaxies, and have been detected 
in all types of galaxies as well as in the intracluster space. Stars in this mass range 
constitute a major component (by mass) of the stellar material in the Universe, 
thus a correct understanding of their evolution in different environments has 
the potential to advance many astrophysical fields. Furthermore, these stars play a 
fundamental role in cosmic recycling, being major contributors to the carbon and
nitrogen 
abundances for the next generation of stars. Stars in this mass range go through the 
asymptotic giant branch (AGB) phase, and through their evolution they enrich the
interstellar medium with helium, carbon, nitrogen, and other elements. 
Typically the abundances of
oxygen, argon, and neon are unaffected during AGB evolution in most stars, thus these 
elements can be used to probe the environment at the time of progenitor formation.

The AGB phase is characterized by the periodicity of nuclear burning phases and dredge-up 
phases, and, as a result, the
products of stellar evolution are carried to the stellar outer layers. 
Planetary nebulae (PN) are the stellar envelopes of AGB stars, ejected at the 
tip of the AGB, carrying the products of stellar evolution to the interstellar medium. 
Thus by analyzing PN abundances one can constraint the AGB evolutionary models, 
and the environment at the time of the formation of the AGB progenitors.
Since the models of stellar evolution predict that the chemical yields form AGB stars 
depend on initial stellar mass and metallicity, large samples of PN and in different
metallicity environments should be compared with evolutionary models to obtain the best 
constraints.
Since PN are ubiquitous and bright, and easily recognizable due to their 
unique spectra, they are ideally suited to probe AGB evolution and populations in 
nearby galaxies. In this paper
we present the state of the art of planetary nebulae abundance analysis in the
Galaxy and the Magellanic Clouds, in particular
the impact of model constraining through carbon abundance, as derived from ultraviolet
spectroscopy. We review large and homogeneous data sets that we used
for our analysis, both for the Galaxy ($\S$2) and the Magellanic Clouds ($\S$3). 
We also present the comparison with the corresponding 
stellar evolution models. A look at the galactic disk population 
and the gradients of elements in the galactic disk is
shown in $\S$4. In $\S$5 we give a summary of the results, and describe the future
challenges of this field.

\begin{figure}[ht!]
\centerline{\includegraphics[width=28pc]{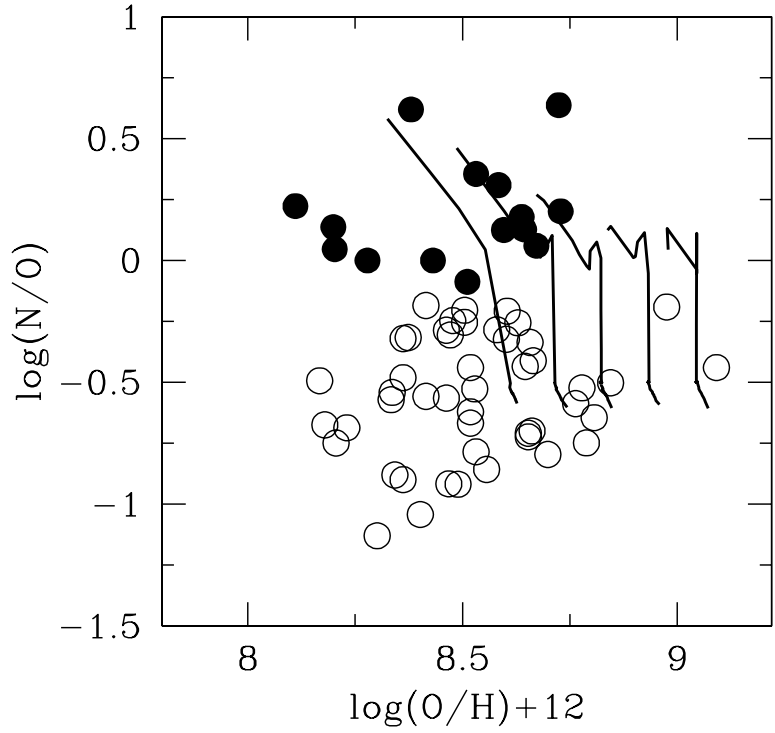}}
\caption{Log (N/O) versus log (O/H) for galactic disk PN. 
Open symbols: non type I PN; filled symbols: type I PN. Lines:
models [7] for z=0.0126, 0.0159, 0.02, 0.0252, and 0.00317 respectively from
right to left.}
\end{figure}

\section{Galactic planetary nebulae}

Abundances of planetary nebulae in the Milky Way disk and bulge have been 
derived by many Authors. Here we focus on large, homogeneous data sets
that have been published recently ([1], [2], [3]. and [4]). Individual
abundances calculated by
these Authors may be different by as much as 20 $\%$, but the differences can be ascribed,
for the most part, to
different ionization correction factors (ICF). Averages of elemental
abundances are also very different across these papers, for the reason that 
some Authors include bulge PN in the galactic averages, while others [4] distinguish
the bulge
contaminants from the disk PN analyzed. For example, the average oxygen abundance for
the whole samples presented in papers [2], [3], and [4] are 
respectively 5.6$\times$10$^{-4}$, 4.3$\times$10$^{-4}$, and 3.5$\times$10$^{-4}$ 
(in terms of O/H, by number). If 
we chose to compare only the common PN the averages
converge respectively to 3.7$\times$10$^{-4}$, 3.9$\times$10$^{-4}$, 
and 3.5$\times$10$^{-4}$, with ranges of the order of $\sigma \sim 2\times10^{-4}$.
The bulge PN have been statistically analyzed by Exter and collaborators [5].
In the following analysis we privilege the sample of papers [4] and [5], respectively
for the galactic disk and bulge, using the data of paper [4] for the disk 
since it excludes explicitly
the bulge population, and it is homogeneously chosen from the IAC morphological catalog of
PN [6]. For carbon abundances in galactic PN we use selected data from [2], and references 
therein.

\begin{figure}[ht!]

\centerline{\includegraphics[width=28pc]{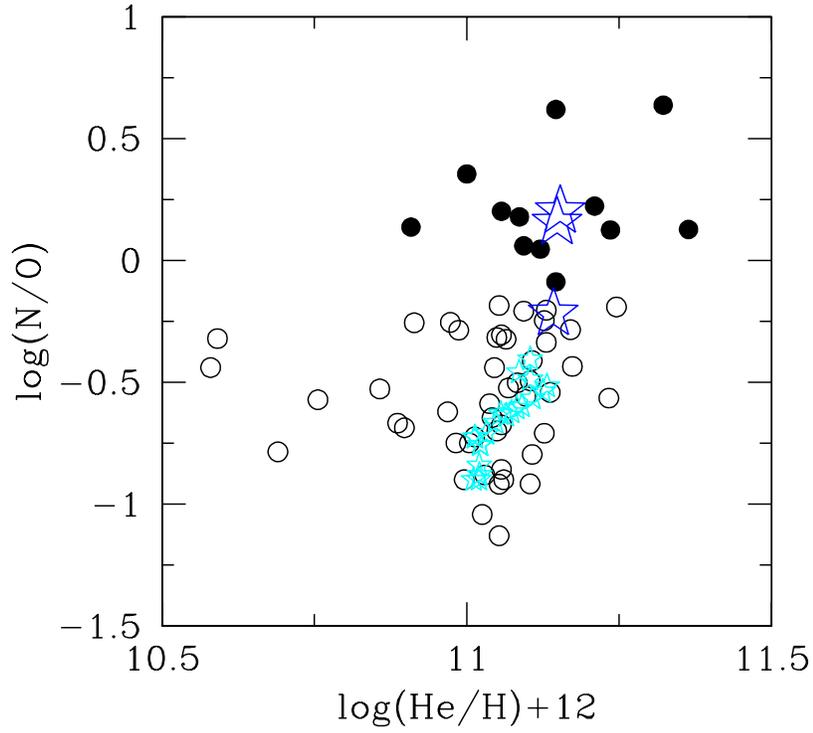}}
\caption{The relation between N/O versus He/H, for galactic disk PN (data from [4]). Open circles: 
non type I PN; filled circles: type I PN; stars indicate models [8] with
M$_{\rm MS}< 4.5$M$_{\odot}$; large stars with 
M$_{\rm MS}> 4.5$M$_{\odot}$.}

\end{figure}

In Table 1, columns 3 and 4, we give the average abundances of key elements, in terms
of the hydrogen abundance, for the 
galactic bulge and the disk PN. In parenthesis we give the range of
the selected diagnostic. We define as type I PN those nebulae 
whose N/O has been enriched with respect to the Orion nebula. 
Determination of carbon abundances for bulge PN are sparse, thus we do not report their average. 
The data show that the oxygen abundance is higher in bulge than in disk PNe, with an 
offset of about 0.1 dex, while the difference in helium abundance is much lower. 
Nitrogen is also more abundant on average in the bulge, making the N/O ratio almost 
identical in the two environments.

In Figure 1 we show the N/O versus O/H relation for all the galactic disk PN analyzed in
[4]. This plot is a good diagnostics to check weather the nitrogen in AGB stars is produced 
by ON cycle. In this case, oxygen would be destroyed to produce nitrogen, and a
strong anti-correlation should be detected between these two ratios. The plot 
shows a large spread of oxygen abundances, as expected in the galactic disk, and
the models by [7] for M$_{\rm MS} < 8$ M$_{\odot}$ lead the eye to show that oxygen 
depletion occurs as nitrogen is produced,
especially in type I PN.

In Figure 2 we show the diagnostic evolutionary plot of N/O versus He/H again for galactic
disk PN, 
showing very clearly that the type I PN correspond to the locus of the 
high mass models (M$_{\rm MS} > 4.5 $ M$_{\odot}$ in Marigo's [8] models).
The C/O versus N/O diagram for galactic disk PN would be sparsely populated.
Carbon abundances derive from UV lines of IUE spectra, and more recent 
carbon determination for disk PN are unavailable.

\section{Magellanic Cloud planetary nebulae}
Abundances of the most common elements except carbon have been 
acquired over the past by several Authors ([9], [10], [11], [12]).
On the other hand, only a handful of carbon abundance
determinations were available [13] until recently. Together with my collaborators 
we started a series of {\it HST} observations
aimed at expand the dataset of carbon abundances in LMC and SMC PN.
For the LMC, among about 350 PN known [14] only 20 IUE spectra are available, with
only 10 sound carbon abundance determinations [13]. We increased this sample by
observing 24 LMC PN with STIS [15]. The SMC is also poorly sampled, with only 
12 IUE spectra for 60 SMC PN known [16]. We have obtained 12 ACS prism spectra 
and were able to determine the carbon lines. Our work on SMC PN carbon
determination is in preparation. The STIS UV spectra of LMC PN show that the volume 
where the carbon emission
originates is the same that the volume where the hydrogen recombination lines and
the oxygen forbidden lines originate as well [15], and the morphology of the
nebulae is the same through the UV and the optical lines.  

The references above were used to obtain the averages of the elemental abundances
of Magellanic Cloud PN in Table 1, columns (5) and (6). The carbon abundances for the 
LMC includes the samples published in [13] and [15], while the SMC carbon
average only includes the SMC PN in [13]. 

\begin{table}[ht!]
\caption{Average abundances}
\smallskip
\begin{center}
{\small
\begin{tabular}{llllll}
\hline
\noalign{\smallskip}
& & Bulge& Disk& LMC& SMC\\
\noalign{\smallskip}
\hline
\noalign{\smallskip}
He/H& 	whole sample& 	0.11 (0.03)& 0.12 (0.04)& 0.10 (0.03)& 0.09 (0.02)\\
    & 	  type I&       0.12& 0.15& 0.11& 0.09\\
    &    non type I&    0.11& 0.11& 0.10& 0.09\\
    
    &&&&&\\
C/H [10$^4$]&   whole sample &    $\dots$  & 5.7 (6.5)& 3.3 (3.5)& 4.3 (2.5)\\    
&&&&&\\
N/H [10$^4$]&  whole sample   &     2.7 (2.3)& 2.4 (3.5)& 0.94 (0.94)& 0.41 (0.31)\\
&&&&&\\
O/H [10$^4$]&   whole sample  &     4.6 (1.2)& 3.5 (2.0)& 2.1 (1.1)& 0.99 (0.84)\\
& type I&                4.3& 3.1& 1.8& 0.48\\
& non type I&             4.7& 3.7& 2.4& 1.2\\
&&&&&\\
N/O& 	whole sample&	0.68& 0.66& 0.66& 1.1\\
&type I&   1.1& 1.9& 1.3& 2.9\\
&non type I& 0.35& 0.32& 0.22& 0.13\\

\noalign{\smallskip}
\hline
\end{tabular}
}

\end{center}
\end{table}

By examining Table 1 we see that the oxygen abundance is, on average, decreasing 
from the galaxy to the LMC and the SMC, as expected. We continue seeing the oxygen
to be depleted in type I PN, as possible consequence of the ON cycle activity. 
We also notice low helium abundances in LMC and, to a larger extent, SMC PN. 
Finally, carbon production is less efficient, on average, in the Magellanic Cloud
PN with respect to the galaxy. We should note that preliminary analysis of our ACS prism
spectra seem to indicate that carbon production in SMC PN is similar, and not 
larger, than that of LMC PN.

In the two upper panels of Figure 3 we show the N/O versus O/H plot for Magellanic Cloud PN. 
It is remarkable how the oxygen gets depleted in nitrogen rich type I SMC PN, while
the effect is much milder in the LMC PN. It looks like whatever the mechanism at work, 
possibly the ON cycle, it is much more efficient at low metallicities than predicted by Marigo's 
models [8].

By examining the carbon and nitrogen production in LMC and SMC PN (lower panels of Figure 3)
we note that
in both galaxies the effect of carbon depletion in favor of nitrogen enrichment is
noticeable. The models by [8] for the LMC and the SMC
PN are also shown in the figure. We note that the low mass models well
encompass the non type I PN locus, while for higher masses the models are not
adequate. Some other mechanisms must be at work that deplete the nitrogen 
(or better the N/O ratio) in type I PN.

\begin{figure}[ht!]
\centerline{\includegraphics[width=28pc]{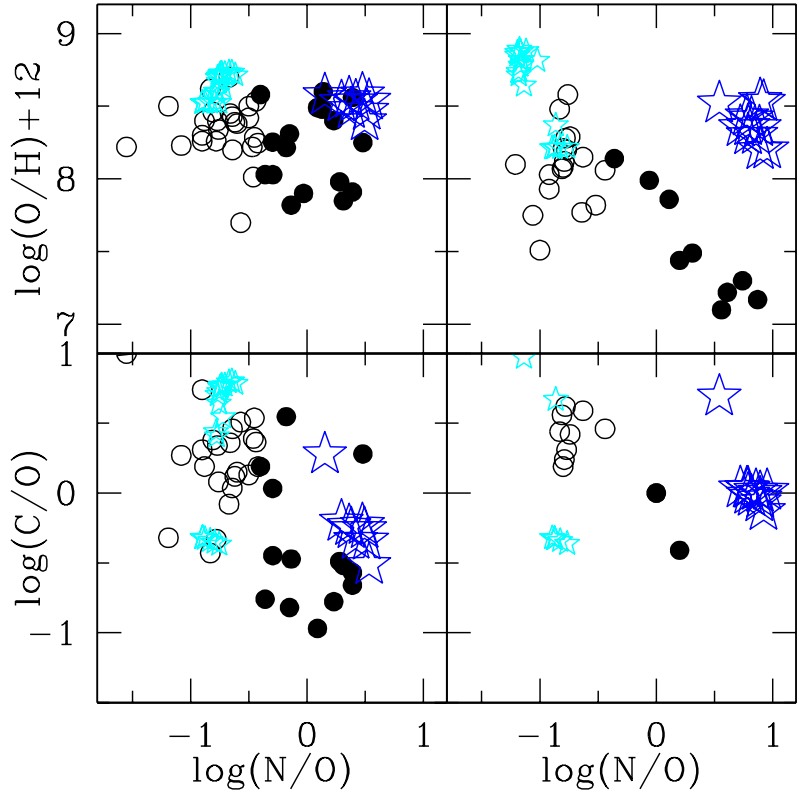}}
\caption{Diagnostic diagrams for LMC (left panels) or SMC (right panels) PN.
Small stars are 
$M_{\rm MS}<3.5$ (LMC) or 3.0 M$_{\odot}$(SMC) models from [8]; 
Large stars represents models with 
$M_{\rm MS}>3.5$ (LMC) or 3.0 M$_{\odot}$(SMC).}
\end{figure}

\section{Probing stellar populations}

Another very important aspect of abundance analysis in planetary nebulae is the study
of the elements that are not affected (or mildly affected) by stellar evolution. 
Recent models (e.g., [7], [8]) predict that at galactic metallicities
oxygen, neon, and argon should not vary considerably during the evolution of these
stars through the AGB, at least in the lower mass range. 
These elements are thus probes of the galactic chemical evolution,
and of the environment at the epoch of progenitor formation. We use here the sample 
of abundances derived in [4] to explore the relations between these elements, 
and to determine galactic gradients and their evolution though time.

We used the oxygen and the neon abundances of PN to determine the metalicity gradients
in the galactic disk. For this study it is essential to exclude bulge PN from the 
samples, since bulge PN may tip the gradients given their average higher metallicity, as
seen in table 1. The gradients from the sample of [4] are shown in Figure 4. We derived
gradients of the order of -0.01 dex kpc$^{-1}$ for oxygen and neon. These gradients agree
with other recent independent determinations of galactic gradients from PN [17], 
as long as the PN sample adopted did not include bulge contaminants. Given that the PN 
gradients are rather flat, 
and that PN represent the progeny of evolved
stars, this picture does not reconcile with flattening of galactic metallicity gradients 
with time. 

Since models of AGB evolution show that nitrogen is enriched especially in high mass AGB
stars via ON cycle, we can use the nitrogen abundance to select 
a more massive sample of PN, and to show the differences in gradients trough the
galactic disk. We found no noticeable differences between gradients of type I and non
type I PN, indicating an overall invariance of gradients with time.

\section{Conclusions and future challenges}

Stellar evolution of low- and intermediate-mass stars is well probed by studying AGB stars
and their remnants. In this paper we have shown how diagnostic abundance plots of PN can be used
to constraint stellar evolution through the AGB. 
Carbon abundances offer a further challenge, since the carbon emission lines 
correspond to the UV part of the spectra. The Magellanic Cloud PN 
allow the comparison with stellar evolution models at lower metallicities.
We have shown that the bulge oxygen and nitrogen abundances are higher, on average, than those of the
galactic disk, suggesting the existence of two distinct stellar populations. We have 
also shown that
the galactic disk population presents a very large spread in environment metallicity (or
metallicity at the formation of the progenitors), and that galactic disk PN
have an effective ON cycle, especially at low metallicity.
This result is somehow strengthened by the analysis of LMC and SMC PN, where the
destruction of oxygen in favor of nitrogen appears clear for SMC PN of type I.
The N/O versus C/O (and versus He/H) diagnostic plots indicate that the type I PN are
the remnant of the evolution of more massive AGB stars.
For the Magellanic Cloud PN the diagrams are sparsely populated, and new data of carbon
abundances are essential for model constraining at low metallicity. 
Galactic gradients of oxygen and neon have been found to be shallow, 
in agreement with other recent and independent PN gradient studies. 

The major future challenge in the field of AGB and post-AGB evolution is twofold: on
the theoretical side, models should take into account the fraction of AGB stars that are
within a binary system, and account for yields of binary evolution; on the observational
side, the acquisition of larger, homogeneous sets of UV spectra to determine carbon and
neon abundances in galactic and local group PN would be essential for progress in this field.
The study of gradients could be extended also to other local group galaxies. Finally, 
PN have been observed also outside galaxies, in the intracluster medium. A better
characterization
of these extragalactic sources would be essential to constraint the mass of dark halos
in galaxies [18], and to determine whether the evolution of these peculiar PN is different 
[19] from that of
galactic PN.

\begin{figure}[ht!]
\centerline{\includegraphics[width=28pc]{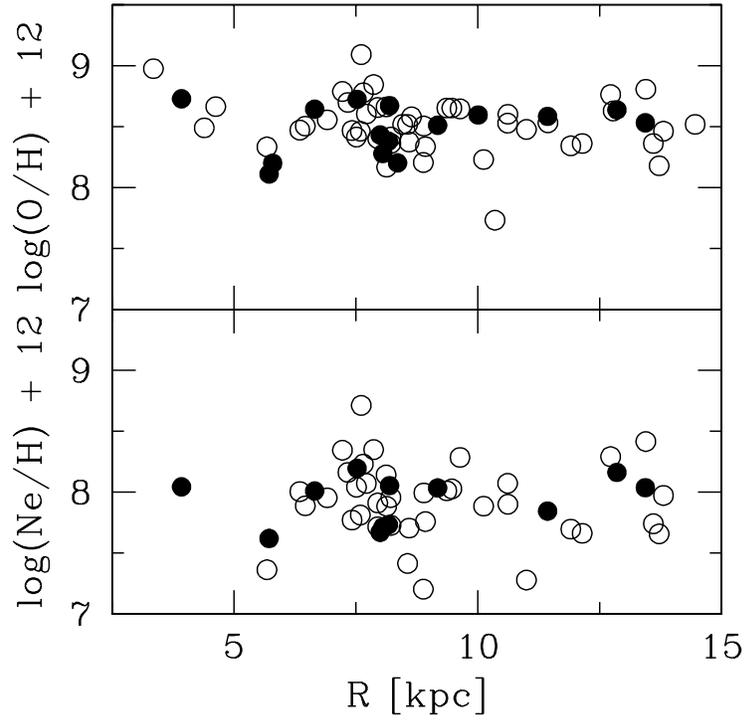}}
\caption{Galactic disk PN metallicity gradients, for oxygen and neon.}
\end{figure}

\noindent
{\it Acknowledgements: Thanks to the Organizers for a very interesting meeting, and to
Katia Cunha, Eva Villaver, Bruce Balick and Dick Shaw for their
contribution to the science presented here.}

\end{document}